# Measurements of the Canonical Helicity of a Gyrating Kink

Jens von der Linden[*] and Jason Sears[†]
*Lawrence Livermore National Laboratory, Livermore, California 94550, USA*

Thomas Intrator[‡]
*Los Alamos National Laboratory, Los Alamos, New Mexico 87545, USA*

Setthivoine You[§]
*Graduate School of Frontier Sciences, University of Tokyo, Tokyo 113-0032, Japan*

(Received 14 March 2018; published 17 July 2018)

Conversions between magnetic and kinetic energy occur over a range of plasma scales in astrophysical and solar dynamos and reconnection in the solar corona and the laboratory. Canonical flux tubes reconcile all plasma regimes with concepts of twists, writhes, and linkages. We present measurements of canonical flux tubes, their helicity, and their helicity transport in a gyrating plasma kink. The helicity gauge is removed with general techniques valid even if only a limited section of the plasma is measured. Temporal asymmetries in the helicities confirm irreducible 3D fields in the kink.

DOI: 10.1103/PhysRevLett.121.035001

A bundle of field lines running through a closed contour form a flux tube. If the field lines are twisted, they form a flux rope. Magnetic flux tubes and ropes are exemplified in coronal loops [1], solar spicules [2], astrophysical jets [3], and discrete temperature flux tubes [4] in tokamak experiments. In ideal magnetohydrodynamics (MHD), magnetic flux is frozen into the plasma. In addition, it has long been recognized that plasmas relax to states of minimum energy subject to the constraint of conservation of magnetic helicity [5], which quantifies the twist, writhe, and interlinking of magnetic flux tubes [6]. Studies suggest that there may also be minimum energy states subject to the conservation of fluid helicity [7] and canonical helicity [8–11], the weighted sum of helicities of magnetic and flow vorticity flux tubes. The dynamics of these relaxation processes can develop microscales, e.g., when flux tubes undergo instability cascades [12–14]. At these microscales, ion inertia, kinetic distribution functions, and finite Larmor radius effects become important, allowing ions and electrons to separate from the magnetic flux, which limits the usefulness of magnetic flux tubes for understanding relaxation dynamics. References [3,15,16] showed that at scales where nonideal effects become important the frozen-in concept can be generalized: regardless of scale each plasma species is frozen-in to the canonical flux related to its canonical momentum. Analytical and numerical studies of collisionless (Hamiltonian) reconnection show that while magnetic flux tubes break during reconnection electron canonical flux tubes stretch but remain continuous [17,18]. A Lagrangian-Hamiltonian formulation can be extended to a field theory for canonical momentum fields obeying Maxwell-like equations [19,20]. The twistedness of the flux tubes is quantified by their helicity, e.g., magnetic helicity for magnetic flux tubes and canonical helicity for canonical flux tubes. Conversions between magnetic and kinetic energy can be viewed topologically as conversions between twisted magnetic and flow flux while conserving the overall helicity of canonical flux tubes of all species [21]. In neutral fluids, e.g., water, it is possible to track vorticity flux tubes with gas bubbles [22] and measure their helicity [23]. Plasmas allow no such tracers, and measuring canonical flux tubes and their helicity requires a volumetric set of measurements. Direct measurements of canonical helicity require either calculating the helicity flux into a volume from surface integrals or integrals of volumetric data [24]. The helicity integrals have a gauge ambiguity that must be removed with reference fields. Magnetic helicity has been indirectly measured in plasmas by reconstructing internal magnetic field profiles with equilibrium models [25] and field line twist numbers [26]. References [27,28] measured magnetic helicity and canonical helicity in reconnecting current sheets and merging flux ropes. References [16,27,28] were able to use the applied fields as reference fields since the measurements spanned the entire vacuum chamber. Here, we report experimental measurements of canonical flux tubes and their helicity evolution and transport in a section of a gyrating kink quasiequilibrium. We apply for the first time techniques for determining reference fields that are valid in any arbitrary experimental volume even if only a limited section of the plasma is measured.

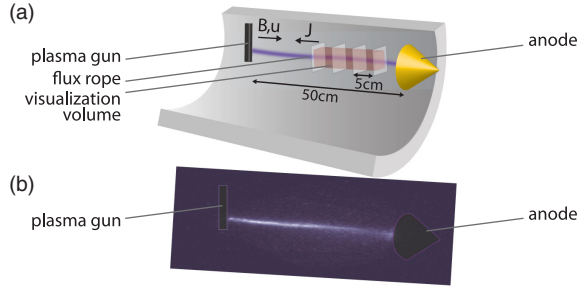

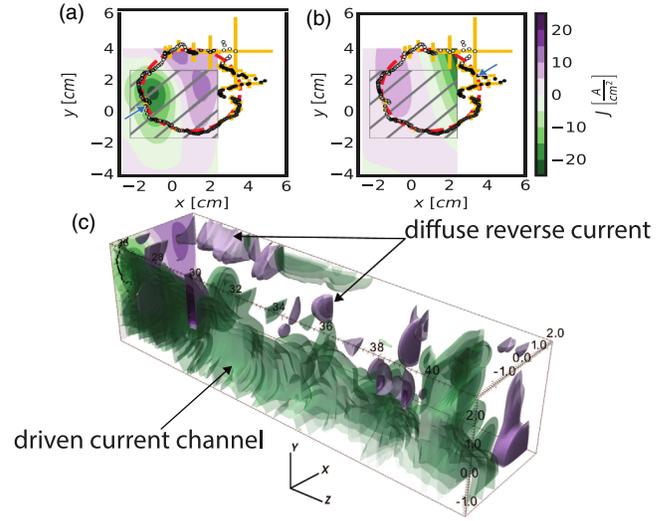

FIG. 1. (a) Schematic of the RSX experiment. The flux rope (purple) is produced by a plasma gun 50 cm away from the anode. Axial flow and applied magnetic field are directed towards the anode. The current flows in the opposite direction and is chosen so that the flux rope is kink unstable, displaces helically, and gyrates. Measurement planes (Fig. 2) and visualization volume (Fig. 3) are drawn in pink. (b) Visible emission from the RSX experiment together with schematics of plasma gun and anode.

This study analyzes 2369 hydrogen shots of a gyrating, kinked flux rope in the reconnection scaling experiment (RSX) [29] (Fig. 1). The RSX generates current-carrying flux ropes of radius $a = 2$–$3$ cm in a cylindrical vacuum chamber of 0.2 m radius between a negatively biased plasma gun and a conical anode fixed at a distance of $L = 0.52$ m from the gun [30] with a uniform axial bias field $B_z = 0.02$ T. As the current ramps up in the flux rope, it kinks, displaces helically, and gyrates. Ohmic heating achieves electron temperatures of 10–15 eV and ion temperatures of 1 eV. The plasma gun achieves an ion to neutral particle ratio of 10 with an ion particle density of $1$–$3 \times 10^{19}$ m$^{-3}$. The plasma gun injects axial flow ($\leq 2.5 \times 10^4$ m/s), directed towards the anode, into the flux rope. This axial flow makes the helical displacement of the kink appear to be gyrating for stationary observers, such as internal probes [31]. Internal Mach, $\dot{B}$, and triple probes measure ion velocity, change in magnetic field, and density, temperature, and plasma potential [32], respectively. All probes are positioned [33] to cover $x$–$y$ planes. The triple and $\dot{B}$ probes cover four axial locations: 24.9, 30.2, 35.7, and 41.6 cm, with the origin at the gun orifice. The Mach probes only measure the Mach number in the $y$ and $z$ direction in the $z = 41.6$ cm plane. Assuming rigid rotation, the Mach numbers in the $x$ direction are the Mach numbers in the $y$ direction shifted by a quarter gyration period. All probe measurements overlap over a cross section of $4.6 \times 4.3$ cm. The canonical flux tubes are plotted in a visualization volume formed by the cross section and the axial extent of the measurements. The helicity is calculated from integrals of the visualization volume with 4%–7% of each side removed to account for truncation errors in the discrete cosine transforms required for the reference field calculations. Since Mach measurements are only available in the fourth plane, they are assumed to be axially uniform. The axial variation of the ion velocity stems solely from the variation of the electron temperature. A spatial resolution of 3 mm is achieved, resolving the ion inertial length (4 cm) and within the same order of magnitude as the thermal ion Larmor radius (5 mm). The digitization rate of 20 MHz resolves the ion gyrofrequency (300 kHz). Reference [30] found that the shape of the kinked flux rope is unchanging while it gyrates. Asymmetries in the plasma current channel and the appearance of a reverse "eddy" current in the $z = 24.9$ cm plane, which does not form a current channel along the $z$ axis, make the system irreducibly 3D (Fig. 2). The irreducible 3D and non-MHD nature lends itself to consider canonical flux tubes; so here, we visualize the 3D evolution of the canonical flux tubes and calculate their helicity. Canonical flux tubes evolve so that their cross sections always enclose a constant flux of canonical circulation, the curl of canonical momentum

$$\vec{\Omega}_\sigma = \vec{\nabla} \times \vec{P}_\sigma$$
$$= n_\sigma m_\sigma \vec{\omega}_\sigma + n_\sigma q_\sigma \vec{B} + \vec{\nabla} n_\sigma \times (m_\sigma \vec{u}_\sigma + q_\sigma \vec{A}), \quad (1)$$

where $\sigma$ denotes the plasma species (e.g., $e$ for electrons and $i$ for ions), $q_\sigma$, $m_\sigma$, and $n_\sigma$ are the species' charge, mass, and density, $\vec{u}_\sigma$ is the species flow, $\vec{\omega}_\sigma = \vec{\nabla} \times \vec{u}_\sigma$ is the species flow vorticity, and $\vec{B} = \vec{\nabla} \times \vec{A}$ is the magnetic field, curl of magnetic vector potential. In a plasma with nonuniform density, both canonical circulation and helicity must be weighted by density. The density gradient terms can be ignored for the instantaneous helicity but must be included in helicity transport calculations. Relative canonical helicity $K_{\sigma\mathrm{rel}}$ quantifies the interlinkings, twists, and writhes of canonical flux tubes and can be expressed as sum of three helicities [34]

$$K_{\sigma\mathrm{rel}} = \mathcal{H}_{\sigma\mathrm{rel}} + \mathcal{X}_{\sigma\mathrm{rel}} + \mathcal{K}_{\sigma\mathrm{rel}}, \quad (2)$$

where magnetic helicity $\mathcal{K}_{\sigma\mathrm{rel}} = q_\sigma^2 \int n_\sigma^2 \vec{A}_- \cdot \vec{B}_+ dV$ quantifies the twistedness of magnetic flux tubes, cross helicity $\mathcal{X}_{\sigma\mathrm{rel}} = m_\sigma q_\sigma \int n_\sigma^2 (\vec{u}_{\sigma-} \cdot \vec{B}_+ + \vec{B}_- \cdot \vec{u}_{\sigma+}) dV$ quantifies interlinkings between magnetic and flow vorticity flux tubes, and flow helicity $\mathcal{H}_{\sigma\mathrm{rel}} = m_\sigma^2 \int n_\sigma^2 \vec{u}_{\sigma-} \cdot \vec{\omega}_{\sigma+} dV$ quantifies the twistedness of flow vorticity flux tubes. The $\pm$ subscripts offset a vector $\vec{X}$ by a reference field $\vec{X}_\pm = \vec{X} \pm \vec{X}_{\mathrm{ref}}$, which is necessary to remove the gauge dependence of helicity [21]. The normal component of the reference fields $\vec{B}_{\mathrm{ref}}, \vec{\omega}_{\mathrm{ref}}$ must be equal and opposite to the physical fields at the boundaries [3]. The transport of relative canonical helicity in a static volume is the sum of four terms [35]

$$\frac{dK_{\sigma\mathrm{rel}}}{dt} = \mathcal{S}_\sigma + \mathcal{A}_{\sigma\mathrm{ref}} + \mathcal{D}_\sigma + \mathcal{R}_{\sigma\mathrm{ref}}, \quad (3)$$

where $\mathcal{S}_\sigma = \int h_\sigma \vec{\Omega}_\sigma \cdot d\vec{S}$ represents static injection from an applied enthalpy $h_\sigma$ at the boundaries, $\mathcal{A}_{\sigma\mathrm{ref}} = \int \vec{P}_{\sigma\mathrm{ref}} \times \partial \vec{P}_{\sigma\mathrm{ref}}/\partial t \cdot d\vec{S}$ represents time-dependent injection, $\mathcal{D}_\sigma = \int \vec{\mathbb{E}}_\sigma \cdot \vec{\Omega}_\sigma dV$ represents sink and source terms as alignment of canonical electric field $\vec{\mathbb{E}}_\sigma = \vec{\nabla} h_\sigma + \partial \vec{P}_\sigma/\partial t$ with canonical vorticity, and $\mathcal{R}_{\sigma\mathrm{ref}} = \int \partial(\vec{P}_{\sigma\mathrm{ref}} \cdot \vec{\Omega}_{\sigma\mathrm{ref}})/\partial t dV$ is the contribution due to changes in reference fields. The enthalpy $h_\sigma = n_\sigma(q_\sigma \phi + k_B T_\sigma + 1/2 m_\sigma u_\sigma^2)$ combines the electrostatic potential $\phi$, the pressure, and the energy of the bulk species' flows, where $k_B$ is the Boltzmann constant. Equation (3) is equivalent to Eq. (12) in Ref. [21] but makes the contribution due to changes in the reference fields, e.g., changes in perpendicular components of boundary piercing fields during flux rope motion, explicit. In the RSX, relative electron canonical flux reduces to magnetic flux, weighted by density, because the product of electron flow and vorticity is negligible compared to the ratio of electron charge over mass. In this study, we consider the relative magnetic, relative ion cross, relative ion flow, and total relative ion canonical helicity and transport.

FIG. 2. Axial current density $j_z$ profiles in the $z = 24.9$ cm plane at (a) $t = 7.34$ $\mu$s and (b) $t = 15.84$ $\mu$s, and $j_z$ isosurfaces in the visualization volume at $t = 7.34$ $\mu$s [36]. While the driven current forms a current channel, the reverse current is diffuse. Here, $j_z$ contours are shown where measurements of $B_x$ and $B_y$ were made. The hatched plane is the $x$–$y$ cross section of the visualization volume. The field null positions over one gyration period of 17 $\mu$s are plotted as dots with time progression corresponding to change in color from white to black and a circle fit (dashed red). Error bars (yellow) are plotted on every 6th field null. A blue arrow marks the field null position at plot time. The isosurfaces (c) enclose $j_z \leq -10$ A/cm$^2$ (green) and $j_z \geq 10$ A/cm$^2$ (purple).

The canonical quantities are calculated from the linear interpolations of measurements. The current is calculated from derivatives of cubic spline fits to the magnetic field measurements. The reference circulation fields [24] are calculated by solving the Laplace problem with discrete cosine transforms [37]. The vector potentials are determined by choosing the DeVore gauge [38], i.e., setting their axial component to zero.

Canonical flux tubes are visualized by integrating field lines of the canonical circulation vector fields. At each time step, a circle of field lines is launched around the field null of the $B_{xy}$ field [39]. The field null is found by successively fitting circles to the curvature of the $B_{xy}$ field and stepping towards the center. Fitting with successively filtered data estimates the error, which is negligible except when extrapolating outside the $B_{xy}$ measurement space. The time evolution of the field null in the $z = 24.9$ cm plane follows a circle centered at $x = 1$ cm and $y = 1.3$ cm, and with radius 2.5 cm (Fig. 2). Figure 3 shows the evolution of two electron and two ion canonical flux tubes. The electron canonical flux tube shape is dominated by the axial bias magnetic field which is an order of magnitude larger than the magnetic field due to internal currents. The reverse current gyrates into the visualization volume as the field null and driven current leave the volume. A single field line wraps clockwise around the flux tube, indicating that canonical electron helicity is negative as expected for an applied current driven antiparallel to the bias magnetic field. The twist of the electron canonical flux tube in the visualization volume switches sign as the field null leaves the volume and the reverse current enters the volume. The peak density and temperature overlap with the electron canonical flux tubes [40]. The ion canonical flux tubes disperse over a wider cross section because the ion vorticity has a larger amount of error due to amplification of noise by the derivatives. The ion canonical flux tubes show significant right-handed twist and writhe around the electron flux tubes, indicating that the total ion canonical helicity is positive as expected for ion flows driven in the same direction as the bias magnetic field.

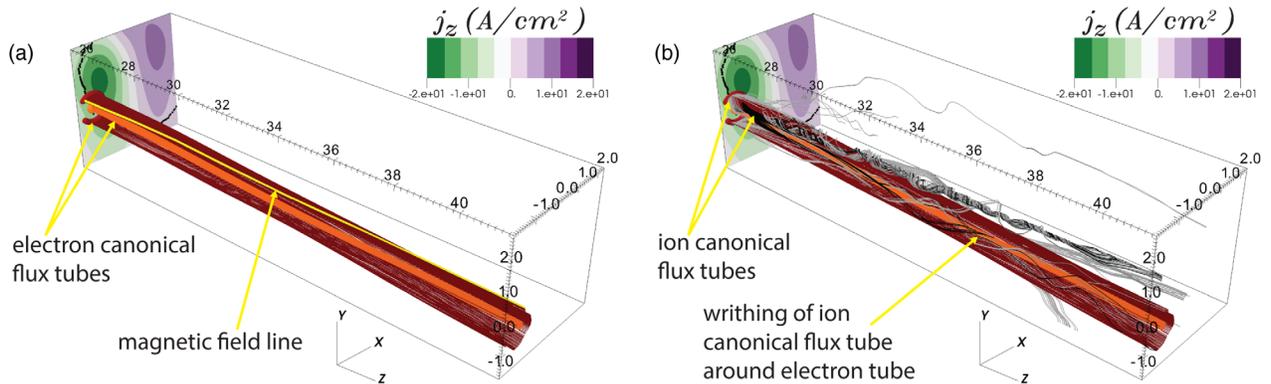

FIG. 3. Gyrating electron (a) and ion canonical (b) flux tubes in the visualization volume shown in Figs. 1 and 2 at $t = 7.34$ $\mu$s [36]. (a) $n\vec{B}$ ($n\vec{\Omega}_e$) field lines trace two electron canonical flux tubes (orange and red). (b) $n\vec{\Omega}_i$ field lines trace two ion canonical flux tubes (black and gray) in the visualization volume. The field lines are launched from circles of 1 mm (orange and black) and 5 mm (red and gray) radii centered at the null of $B_x$ and $B_y$ in the $z = 24.9$ cm plane. The field line null gyration path is shown in black (compare Fig. 2). In (a), a single yellow magnetic field line on the outer flux tube helps visualize the twist of the electron canonical flux tubes. Contours of $j_z$ are plotted in the $z = 24.9$ cm plane.

The strength and sign of the flux tube twist agrees with the evolution of the magnitudes and signs of relative magnetic, relative cross, relative flow helicity, and their sum, which is the total relative ion canonical helicity [Fig. 4(a)]. To provide a reference, the measured helicities are plotted with model helicities of an $a = 2$ cm radius rod with uniform $I = 320$ A current and exponential density cloud, with a peak density of $n_e = 2 \times 10^{19}$ m$^{-3}$ and $e$-folding length of 0.02 m, gyrating along the fitted gyration path from Fig. 2(a) over a 17 $\mu$s period through a $B_0 = 0.02$ T axial bias field and helical flow field with axial velocity $u_z = 25$ km/s and azimuthal velocity $u_\theta = 10$ km/s. While perfect agreement is not expected with a symmetric model, the modeled and measured helicities agree in shape and amplitude. The model helicities peak when the driven current is almost centered in the visualization volume, whereas the measured helicity peaks are delayed by $\sim 2$ $\mu$s. The time integrated helicity transport terms are plotted in Fig. 4(b). The time-dependent injection of helicity is negligible. The static injection of helicity is negative, corresponding to the magnetic helicity injected by the potential difference between gun and anode. The sink and source term is positive, counteracting the static injection through resistive decay. The largest term is the change in reference fields. Since the reference fields are fully determined by the boundary values of the measured fields, this term corresponds to the change in helicity due to the motion of the flux rope in and out of the visualization volume. The total time integrated ion canonical helicity transport roughly matches the shape of the helicity volume integral; however, the time integrated transport is $\sim 10$ times larger than the helicity volume integral. Exact agreement is not expected as the transport terms require taking derivatives in time and space amplifying noise in the measurements.

There is also the possibility of locally increased dissipation and helicity conversions especially since the radius of the current channel is on the order of the ion inertial length (4 cm). A kinetic simulation of a periodic current-carrying magnetic flux tube with similar parameters to the single flux rope in the RSX identified possible reconnection sites [41]. Conversions from magnetic to fluid helicity may occur at these sites. The location of kink driven reconnection in an initially symmetric flux rope may vary over shots in which case the conditional sampling would average out reconnection signatures and the increased helicity dissipation in the transport terms. This is in contrast to flux rope merging experiments where the second flux rope acts as an imposed perturbation making the location of reconnection coherent across shots. The absence of time delay in the model when compared to the measured helicities and integrated transport is due the asymmetric driven current profile and the diffuse reverse current not captured by the symmetric model. The reverse current is in the $+z$ direction, generating a counter-clockwise azimuthal field which together with the positive axial magnetic field gives a positive helicity. This positive helicity counteracts the negative helicity from the driven axial current.

This work is the first application of techniques for determining relative helicity that is valid in any arbitrary volume of experimental data, even if only a limited section of the plasma is measured. To first order, the measured helicities and their transport correspond to the helicities generated by an axial current channel gyrating in a steady axial background magnetic field and a helical flow. Temporal asymmetries in measured helicities confirm the irreducible 3D field profiles in the gyrating kink. The decreasing cost of high-speed digitizers with high-channel count, as well as space missions probing ion and electron scales with clusters of spacecraft [42] will make the collection of large data sets and the calculation of canonical helicity feasible. The data analysis techniques developed here can be applied to other experiments, and space plasma phenomena, with potential for dynamic conversions between magnetic, cross, and fluid helicities, such as compact toroid merging experiments and reconnection in the magnetopause and magnetotail. Canonical flux tubes provide a topological perspective on reconnection and dynamo problems as the flux tubes do not break [18], even under nonideal conditions, and the conservation of their helicity constrains the energy transfer between magnetic field and particles.

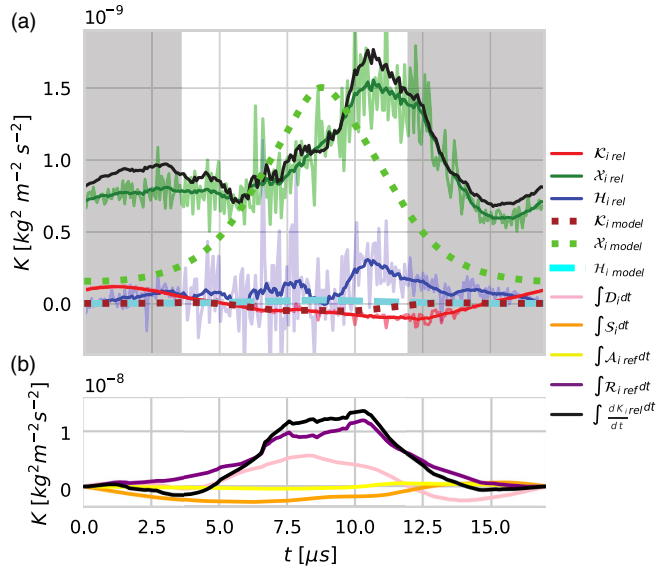

FIG. 4. (a) Measured (solid) and model (dashed) relative magnetic (red), cross (green), flow (blue), and total ion canonical (black) helicity in the visualization volume from Fig. 2. The helicities are boxcar filtered (1.02 $\mu$s width) and normalized to the peak cross helicity. The raw data is plotted with transparency. (b) Time integrated static injection (orange), time-dependent injection (yellow), sources and sinks (pink), change in relative canonical helicity (purple) terms, and their sum, total change in relative canonical ion helicity (black). The mean value of each transport term is subtracted before integration. The white area in (a) marks the time the $z = 24.9$ cm plane field null is inside the visualization volume.

LLNL-JRNL-747364 is based on work supported by a US Department of Energy Grant No. DE-SC0010340, NASA Geospace NNHIOA044I-Basic, and the Center for Magnetic Self Organization funded by NSF and DOE, Office of Fusion Energy Sciences and prepared in part by LLNL under Contract No. DE-AC52-07NA27344. J. v. d. L. acknowledges support by the U.S. Department of Energy, Office of Science Graduate Student Research (SCGSR) program.

*vonderlinden2@llnl.gov
 URL:pls.llnl.gov/people/staff-bios/physics/vonderlinden-j
†sears8@llnl.gov
‡Deceased, June 2014.
§syou@ts.t.u-tokyo.ac.jp